\documentstyle[amssymb,aps,prl,psfig]{revtex}

\begin{document}

\twocolumn[\hsize\textwidth\columnwidth\hsize\csname
@twocolumnfalse\endcsname

\title{Direct observation of the phonon energy in a Bose-Einstein condensate by
tomographic imaging}
\author{Roee Ozeri, Jeff Steinhauer, Nadav Katz and Nir Davidson}
\address{Department of Physics of Complex Systems,\\
Weizmann Institute of Science, Rehovot 76100, Israel}
\maketitle

\begin{abstract}
The momentum and energy of phonons in a Bose-Einstein condensate are
measured directly from a time-of-flight image by computerized tomography. We
find that the same atoms that carry the\ momentum of the excitation also
carry the excitation energy. The measured energy is in agreement with the
Bogoliubov spectrum. Hydrodynamic simulations are performed which confirm
our observation.
\end{abstract}

 ] \bigskip
 In a Bose-Einstein condensate (BEC), the
long wavelength excitations are phonons, characterized by a linear
dispersion relation, as well as a wave function consisting of
atoms with positive and negative momenta. Both the dispersion
relation and the wave function were measured indirectly, by Bragg
spectroscopy \cite{our experiment} \cite{Bogoliubov meas.}.

A two-photon Bragg process coherently imparts an energy $\hbar
\omega $\ and momentum $\hbar k$, determined by the frequency
difference and the angle between the Bragg beams, to the
condensate \cite{Bragg Phillips}. In Bragg spectroscopy, the
two-photon energy $\hbar \omega $ is varied, and the response of
the BEC to the two-photon Bragg transition is observed \cite
{Bragg spectroscopy} \cite{Excitation of Phonons} \cite{our
experiment}. Thus, the directly measured energy is the photon
energy. Calculating the excitation energy requires knowledge of
the process by which the photons impart momentum and energy to the
BEC. Analogously, excitations in superfluid $^{\text{4}} $He were
observed indirectly, by measuring the momenta and the energies of
scattered neutrons \cite{He}.
\begin{figure}[h]
\begin{center}
\mbox{\psfig{figure=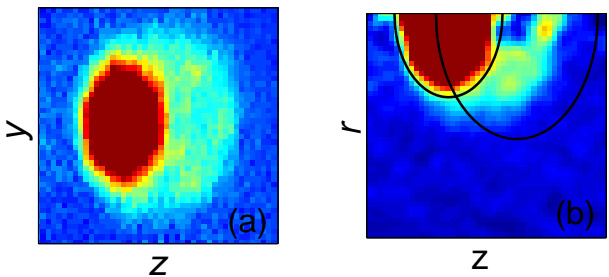,width=8.0cm}}
\end{center}
\vspace{0.4cm} \caption{(a) Absorption picture of a BEC excited
with $k\xi =0.96$ after 38 msec TOF. The left cloud corresponds to
the condensate and the right cloud to the released phonons. (b)
The density distribution of (a) reconstructed by computerized
tomography. A clear separation is evident between the two clouds.
The energy is computed within ellipsoids of sizes indicated by the
ellipses. The area of each image is $400\mu m\times400\mu m$.}
\end{figure}
The phonon energy is much larger than that of a free particle with
the same momentum, and consists mostly of interaction energy
between the phonon and the condensate. When the trapping potential
is turned off, all the interaction energy is transformed into
kinetic energy during a short acceleration period \cite{TOF}.

In this letter we present a direct measurement of this energy from
time-of-flight (TOF) images of the released atoms. Previously the
ground state interaction energy of the condensate was also \
measured from TOF absorption images \cite{TOF}. Since absorption
images provide only the density of the cloud integrated along the
axis of the absorption beam, whereas energy measurements require
knowledge of the full density distribution, fits to model
functions of the density were needed \cite{TOF}. Here we use
computerized tomography of the TOF absorption images, in which the
cylindrical symmetry of the cloud is used to reconstruct the full
density profile from a single absorption image.

Several measurements are performed using this technique. First,
the ground state interaction energy of a released, unexcited BEC
is measured and compared to the value extracted from a fit to the
radial size of the absorption image. Second, the density
distribution of a Bragg-excited condensate is reconstructed. A
clear separation between a released-phonon cloud and a condensate
cloud, which is not clear in the original absorption image, is
visible. The energy and momentum added to the cloud by the Bragg
pulse are measured from the reconstructed image. In this way the
excitation energy is found without sweeping through frequencies,
and by looking directly at images of the atoms. Finally, the
energy carried by the released phonons is measured by observing
the distribution of the total energy between the two clouds. Thus,
we extract the excitation energy from a single absorption image.
Hydrodynamic simulations of the phonon release process are
performed and found to be consistent with our observations.

In a previous experiment the interaction energy between two
identical condensates, with a large relative velocity, was
measured by their interference pattern \cite{Phillips}. For this
case a one-dimensional treatment was sufficient.

The relation between $E_{k}$, the energy, and $\hbar k$, the
momentum, of an excitation in a homogenous condensate has the Bogoliubov form $E_{k}=\sqrt{%
E_{r}\left( E_{r}+2gn\right) }$\ \cite{Bogoliubov}, where $n$ is
the condensate density, $g$ is the coupling constant, given by
$g=4\pi \hslash ^{2}a/m$, $m$ is the mass of the atoms, $a$ is the
scattering length, and $E_{r}=\left( \hslash k\right) ^{2}/2m$\ is
the corresponding free-particle energy. According to the Feynman
relation, the probability of creating an excitation with momentum
$\hbar k$ is proportional to $E_{r}/E_{k}
$\ \cite{Feynman}. Excitations with $k<\xi ^{-1}$\ , where $\xi =\sqrt{%
\hslash ^{2}/2mgn}$\ is the healing length, are called phonons
whose dispersion relation is approximated by the linear relation
$E_{k}\thickapprox \sqrt{gn/m}k$ \cite{Pines}.

For an inhomogeneous condensate, we use the local density
approximation (LDA) according to which the system behaves locally
as a part of a homogenous condensate. The excitation energy is
then defined by the average of the local excitation energies over
the entire condensate \cite{Dynamic structure factor},

\begin{equation}  \label{Ek}
E_{k}=\frac{1}{NS_{k}}\int \epsilon _{k}\left( r\right) \frac{E_{r}}{%
\epsilon _{k}\left( r\right) }n\left( r\right) dr
\end{equation}
where $\epsilon _{k}\left( r\right) $ is the excitation energy
corresponding to $n\left( r\right) $, $S_{k}={\displaystyle{1
\over N}} \int {\displaystyle{E_{r} \over \epsilon _{k}\left(
r\right) }} n\left( r\right) dr$ is the structure factor of the
condensate, and $N$ is the number of atoms.

In our experiment a nearly pure BEC of approximately 10$^{\text{5}}$ $^{%
\text{87}}$Rb atoms in the $\left| F,m_{f}\right\rangle =\left|
2,2\right\rangle $\ \ ground state, is formed in a QUIC type
magnetic trap \cite{QUIC}. The thermal component of the cloud is
measured to be 5\% or less. The trap is elongated and
cylindrically symmetric, with radial and axial trapping
frequencies of $220$\ Hz and $25$ Hz, respectively. The radial and
axial Thomas-Fermi (TF) radii of the condensate
are thus $R_{TF}=3$ $\mu m$ and $Z_{TF}=27$ $\mu m$ \cite{Theory of BEC}. $%
\xi ^{-1} =4.3$ $\mu m^{-1}$ averaged over the entire condensate.

The two Bragg beams, produced by the same laser, are detuned 6.5
GHz from the 5S1/2, F=2 $\longrightarrow $ 5P3/2, F'=3 transition.
The beams pass through two acousto-optic modulators that control
the frequency difference between them. Bragg pulses of duration
varying between 1 and 3 msec are applied to the condensate. The
angle between the two beams is varied to produce excitations with
varying momenta along the $z$ axis (to ensure cylindrical
symmetry). The beams intensities are adjusted such that the number
of atoms in the excitation is always less than 25\% of the total
number of atoms in the condensate. The frequency difference
between the two beams is varied to produce excitations of
different strengths. In the range of $k$'s measured, the dominant
broadening of the Bragg resonance is due to the inhomogeneous
density in the condensate \cite{Dynamic structure factor}, and is
always narrower than the finite-time broadening of our pulses.
Since the pulse frequency spectrum is broad compared to the
excitation line width, excitations are formed at every point in
the condensate \cite{Pulse frequency}. Thus, the excitation energy
is given, according to Eq.(1), by the average energy added per
excitation, i.e. the energy added to the cloud by the pulse,
divided by the number of excitations.

After the Bragg pulse, the magnetic trapping potential is rapidly
shut off, and repulsion between atoms causes a short acceleration
period, during which all the interaction energy is transformed
into kinetic energy. After 38 msec of free expansion the cloud is
imaged by an on-resonance absorption beam, perpendicular to $z$.
Figure 1a shows an absorption image for excitations with $k\xi
=0.96$. Two overlapping clouds are seen. The left cloud
corresponds to the condensate, and the right cloud to the released
phonons. The right cloud is clearly larger in the radial direction
than the left cloud, reflecting the interaction energy of the
phonons.

In the TOF image, the atom's position is proportional to their momentum.
Therefore, if the density distribution $n(r,z)$ after free expansion is
known, we can directly calculate the energy and momentum of the atoms. The
absorption picture only provides $n(r,z)$ integrated along the absorption
beam axis $f(y,z)=\int n(r,z)dx$. However, since the cloud is cylindrically
symmetric we reconstruct $n(r,z)$ from a single absorption image using
computerized tomography. According to the Fourier slice theorem, $n(r,z)$ is
reconstructed by evaluating \cite{Tomography}

\begin{equation}  \label{reconstruction}
n(r,z)=\frac{1}{\left( 2\pi \right) ^{2}}\int F\left( \kappa
_{y},z\right) J_{0}\left( \kappa _{y}r\right) d\kappa _{y}
\end{equation}
where $F\left( \kappa _{y},z\right) $ is the 1D Fourier transform
of$\ f(y,z) $ along the y direction, and $J_{0}$ is the zeroth
order Bessel function. The center of coordinates $(r=0,z=0)$ is
determined by a Gaussian fit to the BEC cloud. Figure 1b shows the
reconstructed $n(r,z)$ of Fig. 1a. The released-phonon cloud is
seen to have a shell-like shape. A clear separation between the
two clouds, which is not visible in Fig. 1a, is now seen. This
separation is caused by the released repulsion energy between the
two clouds.

From the reconstructed $n(r,z)$, the energy of the atoms is calculated as,

\begin{equation}  \label{Energy}
E=\frac{\pi m}{\tau ^{2}}\int n\left( r,z\right) \left[ r^{2}+\left(
z-z_{in}\left( z\right) ^{2}\right) \right] rdrdz
\end{equation}
where $\tau =38$ msec is the time of flight, and $z_{in}(z)$ is
the in-trap z position of an atom with position z after TOF
($r_{in}$ is approximately $0 $). For the BEC cloud we take
$z_{in}\left( z\right) =\left( z/Z_{final}\right) Z_{TF}$, where
$Z_{final}$ is the parabolic radius in the $z$ direction of an
integrated parabola, fitted to the absorption image \cite{TOF}.
For the released-phonon cloud, we have no theory for $z_{in}(z)$. Thus, we take $%
z_{in}(z)=0$.

First, the energy of an unperturbed BEC is measured. To reduce
background we measure the BEC energy within an ellipsoid with
radii obtained from a parabolic fit to the absorption image. The
average energy per atom is measured in 12 images of unperturbed BEC's to be $E_{BEC}/h$ $%
=640$ Hz, which is approximately equal to the value of $\ E_{BEC}/h=590$ Hz,
obtained by the more conventional technique of fitting an integrated
parabola to the same absorption images.

The energy of a Bragg-excited BEC is then measured in the combined
volume of two ellipsoids, centered around the BEC cloud, and the
released-phonon cloud (see Fig. 1b). The size of the second
ellipsoid is enlarged to include all the atoms in the
released-phonon cloud. A reference image with no excitations is
subtracted in the area of the released-phonon cloud to exclude the
energy carried by the thermal atoms. The measured momentum of the
cloud, divided by $\hbar k$, yields the number of excitations.
During the pulse the average energy per atom in the condensate
cloud $E_{BEC}$ decreases by $\sim $12\%. We take the time-averaged value $%
E_{BEC}/h=600$ Hz to be the relevant value.
\begin{figure}[h]
\begin{center}
\mbox{\psfig{figure=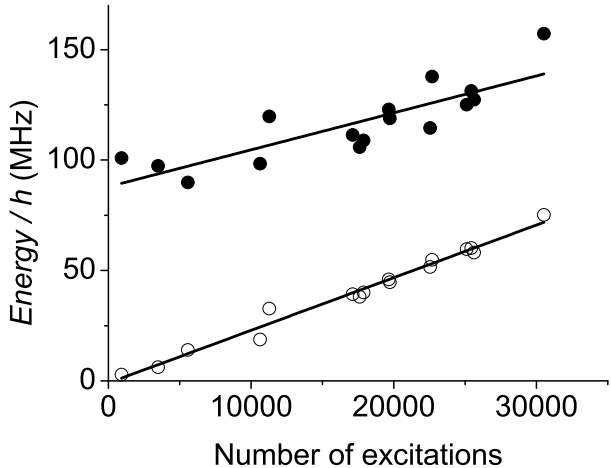,width=8.0cm}}
\end{center}
\vspace{0.4cm} \caption{The filled circles are the measured energy
in the condensate and released-phonon clouds as a function of the
number of excitations for $k\xi =0.96$. The open circles are the
energy measured in the released-phonon cloud only. Solid lines are
the linear fit to the data. }
\end{figure}
\begin{figure}[h]
\begin{center}
\mbox{\psfig{figure=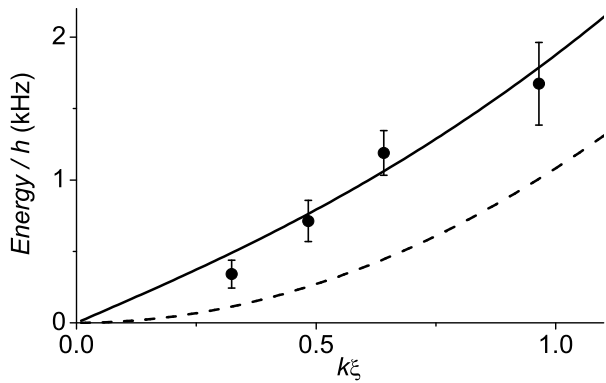,width=8.0 cm}}
\end{center}
\vspace{0.4cm} \caption{Measured excitation energies, for
excitations of various $k$ extracted from the slope of the energy
in both clouds versus the number of excitations. The solid line is
the Bogoliubov dispersion relation, calculated in the LDA for the
measured $E_{BEC}$ with no fit parameters. The dashed line shows
the free-particle parabolic dispersion.}
\end{figure}
The filled circles in Fig. 2 are the measured energy of the atoms
in both clouds as a function of the measured number of excitations
for $k\xi =0.96$. The slope of these points is $E_{k}/h=1675\pm \
290$ Hz and corresponds to the energy of a single excitation. It
is in agreement with $E_{k}/h=1780$ Hz, calculated from the
Bogoliubov dispersion relation in the LDA and also with the
measured value of \cite{our experiment} $E_{k}/h=1750\pm \ 20$ Hz.
Figure 3 shows the excitation energies measured this way at four
different momenta. The measured energies agree well with the
Bogoliubov excitation spectrum for our measured $E_{BEC}$ in the
LDA with no fit parameters, indicated by a solid line.
\begin{figure}[h]
\begin{center}
\mbox{\psfig{figure=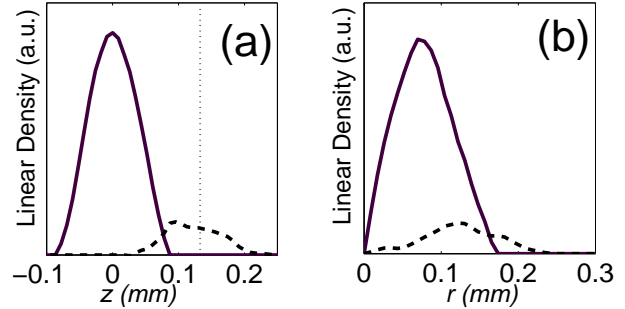,width=8.0cm}}
\end{center}
\vspace{0.4cm} \caption{The linear density of the condensate cloud
(solid line) and the released-phonon cloud (dashed line), for
$k\xi =0.96$ after 38 msec of flight. (a) z direction, (b) r
direction. The plot shown is an average of several images. The
position of the Bragg recoil from the center of the condensate is
indicated by the vertical dotted line. The released-phonon cloud
is clearly larger than the condensate cloud in the radial
direction. The condensate cloud is not distorted by the release
process.}
\end{figure}
In the reconstructed $n(r,z)$ of Fig. 1b, the condensate cloud and
the released-phonon cloud are distinct and separate. Figure 4a and
4b show the linear density profile in each of the clouds, along
the z and the r directions respectively, for $k\xi =0.96$. The
released-phonon cloud is significantly larger than the condensate
cloud in the r direction.

The condensate cloud seems not to be distorted by the interaction
with the phonons. This may imply that each phonon is repelled by a
fraction of the condensate that is much heavier than the phonon.
Since the condensate cloud is much heavier than the phonon cloud,
this would suggest that most of the excitation energy is carried
by the released-phonon cloud. In order to check this hypothesis,
we measure the energy of each cloud separately.\ The open circles
in Fig. 2 are the measured energy in the released-phonon cloud
only, as a function of the measured number of excitations. From a
linear fit, the slope of these points is $2390\pm \ 90$ Hz. When
the average energy of an atom in the condensate cloud is
subtracted from this value \cite{Eavg}, we get an energy of
$1840\pm \ 100$ Hz per excitation in the released-phonon cloud,
which is consistent with the Bogoliubov value $E_{k}/h=1780$ Hz.
This indicates that all of the excitation energy is indeed carried
by the atoms in the released-phonon cloud.

The filled circles in Fig. 5 are the excitation energies, measured
from the slope of the energy of the released-phonon cloud versus
the number of excitations. Except for the $k\xi =0.32$ point, the
measured points agree with the Bogoliubov spectrum in the LDA,
indicated by a solid line.
\begin{figure}[h]
\begin{center}
\mbox{\psfig{figure=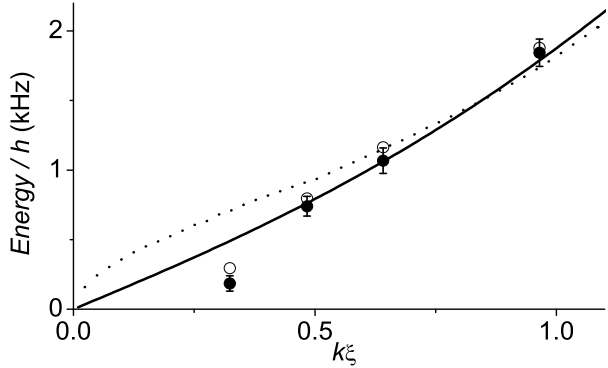,width=8.0cm}}
\end{center}
\vspace{0.4cm} \caption{Excitation energies measured from the
released-phonon cloud only. The filled circles are measured from
the slope of plots like the open circles in Fig. 2. The open
circles are measured from a single image. The solid line is the
Bogoliubov dispersion relation, calculated in the LDA for the
measured $E_{BEC}$ with no fit parameters. The dotted line is the
average energy of the released-phonon cloud minus the condensate
average energy, by a hydrodynamic simulation.}
\end{figure}
Given that all the excitation energy is carried by the
released-phonon cloud, this energy can be extracted from a single
image. This technique requires the collection of far less data
than does the indirect measurement, where an entire spectrum is
needed. The open circles in Fig. 5 are the excitation energies
extracted from the image with the largest number of excitations,
for each $k$. These points agree with the filled circles measured
from the slope.

During the release process some of the interaction energy is
transformed into motion in the $z$ direction, causing the
condensate to recoil. The zero of momentum is therefore not at the
center of the condensate cloud, introducing a systematic error in
our procedure. For the higher $k$ values the interaction energy is
transformed mainly into motion in the radial direction so this
systematic error is negligible. For the smaller $k$ values a
larger fraction of the energy is transformed into motion in the
$z$ direction. This may explain the low energies measured at the
$k\xi =0.32$ point.

We perform hydrodynamic simulations of the release of the excited
cloud \cite{Theory of BEC}. Excitations at different initial
positions in the condensate are chosen with a probability
${\displaystyle{E_{r} \over \epsilon _{k}\left( r\right) }}
n\left( r\right) $ . The condensate is then released and the
excitations propagate with an initial momentum $\hbar k$. The
force at each time step is equal to the gradient of the
interaction energy in the condensate at the current position of
the excitation, multiplied by a factor, accounting for the
excitation's initial interaction energy. Since the excitations are
a small perturbation, the condensate is assumed to expand as would
an unperturbed condensate \cite{TOF}. The calculated average
energy of the excitations minus the average energy of the BEC,
after 38 msec of free expansion, is shown by the dotted line in
Fig. 5. The simulations are seen to be consistent with the
observation that the released-phonon cloud carries all of the
excitation energy.

In conclusion, the interaction energy of phonons in a
Bose-Einstein condensate is directly measured by computerized
tomography of TOF images. The excitation energy is extracted from
a single absorption image. This measurement of the excitation
energy relies solely on images of the atoms, and does not require
any information about the photon energy. Furthermore, no
assumptions are made with respect to the shape of the cloud,
except for cylindrical symmetry. The measured energy is found to
agree with the predicted Bogoliubov spectrum in the LDA. After the
cloud is released, the excitation energy is seen to be carried by
the same atoms that carry the excitation momentum. Hydrodynamic
simulations of the release process are performed, and are found to
be consistent with our experimental findings.

Computerized tomography of cold atomic clouds could also be used
for the observation of non-trivial density profiles, such as the
$s$-wave scattering shell of two colliding clouds, vortices,
solitons, or the fringe pattern of two interfering condensates. In
cases of cylindrical symmetry, computerized tomography requires
only a single absorption image.


\begin{references}
\bibitem{our experiment}  J. Steinhauer et. al., Cond. Mat. 0111438.
\bibitem{Bogoliubov meas.}  J. M. Vogels et. al., Cond. Mat. 0109205.
\bibitem{Bragg Phillips}  M. Kozuma et al., Phys. Rev. Lett. {\bf 82}, 871 (1999).
\bibitem{Bragg spectroscopy}  J. Stenger et. al., Phys. Rev. Lett. {\bf 82}, 4569
(1999).
\bibitem{Excitation of Phonons}  D. M. Stamper-Kurn et. al., Phys. Rev.
Lett. {\bf 83}, 2876 (1999).
\bibitem{He}  D. G. Henshaw and D. B. Woods, Phys. Rev. {\bf 121}, 1266 (1961).
\bibitem{TOF}  Y. Castin and R. Dum, Phys. Rev. Lett. {\bf 77}, 5315 (1996).
\bibitem{Phillips}  J. E. Simsarian et. al., Phys. Rev. Lett. {\bf 85}, 2040 (2000).
\bibitem{Bogoliubov}  N. N. Bogoliubov, J. Phys. (USSR) {\bf 11}, 23 (1947).
\bibitem{Feynman}  R. P. Feynman, Phys. Rev. {\bf 94}, 262 (1954).
\bibitem{Pines}  D. Pines and Ph. Nozieres, ''The theory of quantum
liquids'' (Addison-Wesley, 1966, 1988), Vol. I.
\bibitem{Dynamic structure factor}  F. Zambelli et. al., Phys. Rev. A {\bf 61},
063608 (2000).
\bibitem{QUIC}  T. Esslinger et. al., Phys. Rev. A {\bf 58}, R2664 (1998).
\bibitem{Theory of BEC}  F. Dalfovo et. al., Reviews of Modern Physics {\bf 71},
463 (1999) .
\bibitem{Pulse frequency}  Indeed, no correlation is found between the
energy per excitation of the pulse and the pulse center frequency.
\bibitem{Tomography}  M. Born and E. Wolf, ''Principles of optics''
(Cambridge university press, 7$^{\text{th}}$ edition, 1999) ch.
IV.
\bibitem{Eavg}  For excitations with $k\xi =0.96$, the number of atoms in the released-phonon cloud
is approximately equal to the number of  excitations, and the
condensate cloud energy decreases by roughly $E_{BEC}$ per
excitation. For lower $k$ values the number of atoms in the
released-phonon cloud is less than the number of excitations, and
the ratio between the two is found to be $\sim k\xi $. The
condensate cloud energy however, decreases by roughly $E_{BEC}$
per excitation for all $k$ values.
\end{references}
\end{document}